\documentclass[onecolumn,showpacs,preprintnumbers,amsmath,amssymb]{revtex4}

\usepackage{graphicx}
\usepackage{dcolumn}
\usepackage{bm}

\begin{document}


\title{A Hamiltonian for Relativistic Interacting Many Particles}

\author{Jose A. Magpantay}
\email{jose.magpantay11@gmail.com}
\affiliation{Quezon City 1101, Philippines\\}
\date{\today}

\begin{abstract}
There is no relativistic Hamiltonian for many particles systems  except for free particles and this has been accepted since the 1960s from the works of Currie, Jordan and Sudarshan, Cannon and Jordan, and Leutwyler. The reason for this is that a time translation generator for interacting particles that will satisfy the Poincare algebra seems to be non-existent. Without a Hamiltonian, the usual approach to equilibrium statistical mechanics via the partition function and non-equilibrium statistical mechanics via the Liouville equation to solve for the Gibbs distribution do not exist. This is the problem we will address in this paper, we will derive a relativistic Hamiltonian for many particles with two body interactions. The derivation of the Hamiltonian is hinted by the fact that a system of many charged particles interacting with an electromagnetic field has a Hamiltonian. 

We start then with many relativistic particles interacting with a scalar field. Taken separately, the Poincare generators for the relativistic particles and scalar fields are easily derived and as a matter of fact quite known. They also satisfy the Poincare algebra. Then we will introduce an interaction term for the combined system guided by two conditions (1) the generators for time translation, spatial translation, real rotations and boosts satisfy the Poincare algebra and (2) when we take the non-relativistic limit of the effective action for the relativistic particles after integrating out the scalars, we should get the usual Hamiltonian with two body interactions. However, we find that the problem still persists, the generators of space-time symmetries, which are conserved, still do not satisfy the Poincare algebra. We find that the source of the violation is the built-in time delayed dynamics in relativistic point particle physics. Still we argue that the relativistic Hamiltonian can be used as the starting point in deriving a relativistic thermodynamics from the partition function and a relativistic Gibbs distribution from a relativistic Liouville equation. 
\end{abstract}

\pacs{Valid PACS appear here}
\maketitle

1. A special relativistic treatment of thermodynamics has been in existence since the advent of special relativity. Unfortunately, unlike the transformation of mechanical quantities which are unique, the transformation of temperature is not unique as shown by the following summary - Planck \cite{Planck} and Tolman \cite{Tolman} claimed that a moving body will be cooler, Eddington \cite{Eddington} and Ott \cite{Ott} claimed the opposite (hotter) while Landsberg \cite{Landsberg} and van Kampen \cite{vanKampen} claimed the same temperature as in the rest frame. Einstein on the other hand took all three positions at different times - early he said moving body will be cooler \cite {Einstein}, then he said it will be hotter \cite{Wang} and before he died he thought that the temperature is an invariant with the value given in the rest frame \cite{Liu}. This confusing state of affairs was summarized by Balescu \cite{Balescu} in an equilibrium statistical mechanical analysis where he argued that there is not just three but an infinity of possible self-consistent relativistic thermodynamics that may arise, which he interpreted as some sort of 'gauge freedom'.The three known thermodynamics he summarized follow from specific 'additional criterion', something like specific 'gauge fixing choices'. The use of an analogous gauge principle is confusing because in gauge theories physical quantities are gauge-invariant and does not depend on the choice of gauge, although some calculations are more expedient in a specific gauge. But definitely, the thermodynamic quantities are physical and should have one definite value and not dependent on the choice of gauge. Unfortunately, up till today, what the definite values of these thermodynamic quantities have not yet been settled.

However, Balescu's relativistic equilibrium statistical mechanics has two shortcomings. First, the initial distribution function in a reference frame with speed v ($ tanh s = v $) satisfies a Liouville equation type with s playing the role of time and the system's boost generator K (boost along x axis, for example) taking the role of the Hamiltonian. Why the initial distribution in an arbitrary frame  (with velocity $ tanh s $) should satisfy this type of 'evolution' equation is not clear.  Second, the bigger drawback is that it is already known then that there is no relativistic many particle Hamiltonian for systems with interacting particles. This was already shown by Currie, Jordan and Sudarshan \cite{Currie}for two particles, then shown by Cannon and Jordan for three particles \cite{Cannon} and then shown to be true for any arbitrary number of particles by Leutwyler \cite{Leutwyler}.  Since the boost generator is  expressed in terms of the Hamiltonian (see following discussions), the initial distribution in an arbitrary Lorentz frame is ill-defined. Even the canonical distribution he defines for a system of temperature T and volume V in frame with velocity $ tanh s $, from which he derived the relativistic thermodynamic transformations, also does not exist. Thus, there is no justification for Balescu's starting point because the Hamiltonian does not even exist. 

2. To have a statistical mechanical analysis of relativistic many particle systems, we must have a Hamiltonian. This is precisely what we will derive in this short paper. For a non-interacting, relativistic many particle systems, the relativistic action is
\begin{equation}\label{1}
S_{rp}(\vec{x}^{a}) = \int dt \sum_{a=1}^{N} mc^{2}\left( 1 - \dfrac{1}{c^{2}} \dot{\vec{x_{a}}} \cdot\dot{\vec{x_{a}}} \right)^{\frac{1}{2}}, 
\end{equation}
where for simplicity we assumed the N particles are identical so there is only one rest mass m.
The  generators of the Poincare algebra are easily derived from standard techniques and are given by
\begin{subequations}\label{2}
\begin{gather}
H=\sum_{a=1}^{N}[(\vec{p_{a}})^{2}+m^{2}]^{\frac{1}{2}}, \label{first}\\
P_{i}=\sum_{a=1}^{N}p_{i}^{a}, \label{second}\\
J_{i}=\epsilon_{ijk}\sum_{a=1}^{N} x_{j}^{a}p_{k}^{a}, \label{third}\\
K_{i}=\sum_{a=1}^{N} \left\lbrace \frac{1}{2}{x_{i}^{a}[(\vec{p_{a}})^{2}+m^{2}]^{\frac{1}{2}}+[(\vec{p_{a}})^{2}+m^{2}]^{\frac{1}{2}}x_{i}^{a}}\right\rbrace , \label{fourth}
\end{gather}
\end{subequations}
where H generates time translation, $ \vec{P} $ generates spatial translation, $ \vec{J} $ generates real rotations while $ \vec{K} $ generates boosts. In equation (2b), there is an  implicit sum in $ j, k $. 

The Poincare algebra satisfied by these operators are given by the Poisson bracket relations
\begin{subequations}\label{3}
\begin{gather}
[P_{i},H] = [P_{i},P_{j}] = [J_{i},H] = 0, \label{first}\\
[J_{i},P_{j}] = \epsilon_{ijk}P_{k}, [J_{i},J_{j}] = \epsilon_{ijk}J_{k}, [J_{i},K_{j}] = \epsilon_{ijk}K_{k},\label{second}\\
[K_{i},H] = P_{i}, [K_{i},P_{j}] \delta_{ij}H, [K_{i},K_{j}] = -\epsilon_{ijk}J_{k},\label{third}\\
\end{gather}
\end{subequations}
The above Poincare algebra follows from the fundamental Poisson brackets
\begin{subequations}\label{4}
\begin{gather}
[x_{i}^{a},x_{j}^{b}] = [p_{i}^{a},p_{j}^{b}] = 0, \label{first}\\
[x_{i}^{a},p_{j}^{b}] = \delta_{ij} \delta^{ab}, \label{second}
\end{gather}
\end{subequations}
where $ i, j $ run from $ 1, 2, 3 $ and $ a, b $ run from 1 to N, the total number of particles.

The point of the papers from the early to mid 1960s is that for interacting, relativistic many particle systems, there is no Hamiltonian and from equation (2d), there is also no $ \vec{K} $ so there is no Poincare algebra to guarantee relativistic invariance. We will derive in this paper the expression for the generators for infinitesimal time translation (Hamiltonian),spatial translation (linear momentum), spatial rotations (angular momentum) and Lorentz transformations (boosts). We will show that because of time delay in the effective theory for relativistic particles, the Poincare algebra is not satisfied, i.e., we show the source of the violation of Poincare algebra. But with a relativistic Hamiltonian we can start a statistical mechanical treatment of the relativistic system.

3. The approach we will use is by coupling the relativistic particles to the simplest field theory, a scalar field. The scalar field is defined in the entire 4D Minkowski space-time. Although the relativistic particles may be confined within a finite spatial volume V, its boundary has nothing to do with the scalar field dynamics. The scalar field will couple only with the relativistic particles in a specific way to provide the relativistic two-body potential. 

Let us write down the action for a free scalar field of mass m.
\begin{equation}\label{5}
S_{s}(\Phi) = \int d^{4}x  \left\lbrace \frac{1}{2} \partial_{\mu}\Phi \partial_{\mu}\Phi - \frac{1}{2} m^{2} \Phi^{2} \right\rbrace  .
\end{equation}
From standard field theory techniques, we find that the generators of the Poincare group for this system are given by
\begin{subequations}\label{6}
\begin{gather}
H = \int d^{3}x \textsf{H}(\vec{x},t) ,\label{first}\\
P_{i} = \int d^{3}x \textsf{P}_{i}(\vec{x},t),\label{second}\\
J_{i} = \epsilon_{ijk} \int d^{3}x   \textsf{P}_{j}(\vec{x},t) x_{k} ,\label{third}\\
K_{i} = \int d^{3}x \left\lbrace  x_{i} \textsf{H}(\vec{x},t) + x_{0} \textsf{P}_{i}(\vec{x},t), \right\rbrace \label{fourth}
\end{gather}
\end{subequations}
where the densities in the above formulas are given by
\begin{subequations}\label{7}
\begin{gather}
\textsf{H} = \left\lbrace \frac{1}{2} \Pi^{2}(\vec{x},t) + \frac{1}{2} \nabla\Phi\cdot\nabla\Phi + \frac{1}{2} m^{2} \Phi^{2} \right\rbrace ,\label{first}\\
\textsf{P}_{i}(\vec{x},t) =  \Pi(\vec{x},t) \partial_{i}\Phi(\vec{x},t). \label{second}
\end{gather}
\end{subequations}
Equation (7a) defines the Hamiltonian density, equation (7b) defines the momentum density. 

The Poincare algebra given by equation (3) are easily verified for this free scalar field by making use of the fundamental equal time Poisson brackets given by
\begin{subequations}\label{8}
\begin{gather}
[\Phi(\vec{x},t),\Phi(\vec{x}',t)]= [\Pi(\vec{x},t),\Pi(\vec{x}',t)] = 0,\label{first}\\
[\Phi(\vec{x},t),\Pi(\vec{x}',t)] = \delta^{3}(\vec{x}-\vec{x}'),\label{second}
\end{gather}
\end{subequations}.

For the combined relativistic many (non-interacting), particles and scalar field theory, the action is given by sum of the actions given by equations (1) and (5) with the generators of the Poincare group given by the sums of the generators in equations (2) and (6) and (7). Obviously, they also satisfy the Poincare algebra. 

4. Next we put in the relativistic many particles and scalar field interaction term. The interaction term should be guided by two conditions. First, the combined theory should still respect Poincare symmetry and the generators explicitly satisfy the Poincare algebra. If this is satisfied, we are at least guaranteed that the starting point is Poincare invariant.

Second, when we integrate out the scalar field to get the relativistic effective action for the relativistic many particles and do the non-relativistic limit, $ c \rightarrow \infty $, it must give back the usual kinetic energy term plus two-body instantaneous interaction potential. 

We begin with this consideration. The interaction action we will consider is
\begin{subequations}\label{9}
\begin{gather}
S_{int}(\vec{x}^{a},\Phi) = \int d^{3}x dt \alpha \textsf{j}(\vec{x},t;\vec{x}^{a})\Phi(\vec{x},t),\label{first}\\
\textsf{j}(\vec{x},t;\vec{x}^{a}) = \sum_{a=1}^{N}  \delta^{3}(\vec{x} + \vec{x}^{a}(t)),\label{second}
\end{gather}
\end{subequations}
where $ \alpha $ is the coupling strength of the scalar field and point particles. The plus sign inside the delta function in equation (9b) will become clear when we verify the Poincare algebra. Equation (9) is linear in $ \Phi $ so that when we integrate out $ \Phi $ to derive the effective action for the relativistic particles, we will have an effective relativistic potential, which we can take its non-relativistic limit to give two-body interaction potentials. The effective action for the relativistic particles is given by the path-integral
\begin{equation}\label{10}
\exp {\left( i S_{eff}(\vec{x}^{a})\right)} = \int (d\Phi) \exp {i\left( S_{rp} + S_{s} + S_{int}\right)},
\end{equation}
where the relevant terms are given in equations (1), (5) and (9). The integral is easily done giving
\begin{equation}\label{11}
S_{eff}(\vec{x}^{a}) = S_{rp} -  \frac{1}{2} \int dt dt'\sum_{a,b; a \neq b}^{N} \alpha^{2} G_{4}(\vert\vec{x}^{b}(t') -\vec{x}^{a}(t)\vert; t - t') - \frac{i}{2} tr ln \left(-\nabla^{2} + \dfrac{1}{c^{2}} \partial_{t}^{2} + m^{2} \right) ,
\end{equation}
where $ G_{4} $ is the Green function of the relativistic operator $- \nabla^{2} + \dfrac{1}{c^{2}} \partial_{t}^{2} + m^{2} $. The Fourier decomposition of this Green function is given by
\begin{equation}\label{12}
G_{4}(\vert\vec{x}^{b}(t') -\vec{x}^{a}(t)\vert; t - t') = \int d^{3}k dk_{0} \dfrac{1}{\vec{k}\cdot\vec{k} - \dfrac{k_{0}^{2}}{c^{2}} + m^{2}} \exp {[ik_{0}(t - t') - i \vec{k}\cdot(\vec{x}^{b}(t') - \vec{x}^{a}(t))]}.
\end{equation}

Let us take the non-relativistic limit, effectively we put $ c \rightarrow \infty $ and $ G_{4} $ goes into the instantaneous Yukawa potential
\begin{equation}\label{13}
G_{4}(\vert\vec{x}^{b}(t') -\vec{x}^{a}(t)\vert; t - t') \rightarrow \delta(t - t') \dfrac{1}{\vert\vec{x}^{a} - \vec{x}^{b}\vert} e^ {{-m\vert\vec{x}^{a} - \vec{x}^{b}\vert}}.
\end{equation}

It seems we can only get a specific two-body potential, the Yukawa potential. If we put $ m = 0 $, we will get a Coulomb potential. This does not look promising, we do not get the relativistic potential of an arbitrary two body non-relativistic potential. But there is a solution to this problem. 

5. Suppose we put $ m^{2} = m^{2}(\vert\vec{k}\vert^{2} - \dfrac{k_{0}^{2}}{c^{2}}) $, i.e., the mass of the scalar field  $ \Phi $ dependent on the Lorentz invariant wave number. Note, what is important is the explicit dependence on the invariant $ k_{\mu} $. Different scalars will have different masses but its action is still given by equation (5). Equation (12) will still hold only $ m^{2} = m^{2}(\vert\vec{k}\vert^{2} - \dfrac{k_{0}^{2}}{c^{2}}) $. 

Taking the non-relativistic limit, and if we impose that the resulting instantaneous potential is the two-body non-relativistic potential, then
\begin{subequations}\label{14}
\begin{gather}
G_{4}(\vert\vec{x}^{b}(t') -\vec{x}^{a}(t)\vert; t - t') \rightarrow \delta(t - t') G_{3}(\vert\vec{x}^{a} - \vec{x}^{b}\vert),\label{first}\\
 G_{3}(\vert\vec{x}^{a} - \vec{x}^{b}\vert) = \int d^{3}k \dfrac{\exp {-i\vec{k}\cdot(\vec{x}_{b} - \vec{x}_{a})}}{\vert\vec{k}\vert^{2} + m^{2}(\vert\vec{k}\vert)}.
 \end{gather}
 \end{subequations}
 Equation (14b) can be inverted to give
 \begin{equation}\label{15}
  m^{2}(\vert\vec{k}\vert^{2}) = \left[ \int d^{3}x G_{3}(\vert\vec{x}\vert)\exp {i\vec{k}\cdot\vec{x}} \right] ^{-1} - \vert\vec{k}\vert^{2}.
  \end{equation}
Equation (15) says that the mass spectrum of the scalar field that couples to the relativistic particles is determined by the non-relativistic two-body potential $ G_{3} $. To solve the relativistic two-body potential $ G_{4} $ as given by equation (12) we now must extend the argument of $ m^{2} $ given in equation (15) from $ \vert\vec{k}\vert^{2} $ to  $ \vert\vec{k}\vert^{2} - \dfrac{k_{0}^{2}}{c^{2}} $. By construction this two-body potential is explicitly relativistic, it is determined by the two-body non-relativistic potential and it is obviously not instantaneous. Simplifying, we find
\begin{subequations}\label{16}
\begin{gather}
G_{4}(\vert\vec{x}^{b}(t') -\vec{x}^{a}(t)\vert; t - t') = \int d^{3}k dk_{0} \dfrac{ \exp {[ik_{0}(t - t') - i \vec{k}\cdot(\vec{x}^{b}(t') - \vec{x}^{a}(t))}]}{n^{2}(\vert\vec{k}\vert^{2} - \dfrac{k_{0}^{2}}{c^{2}})},\label{first}\\
n^{2}(\vert\vec{k}\vert^{2}) =  \left[ \int d^{3}x G_{3}(\vert\vec{x}\vert) \exp {i\vec{k}\cdot\vec{x}} \right]^{-1} ,\label{second}
\end{gather}
\end{subequations}
and the argument of $ n^{2} $ in equation (16b) is extended from $ \vert\vec{k}\vert^{2} $ to $ \vert\vec{k}\vert^{2} - \dfrac{k_{0}^{2}}{c^{2}} $ in equation (16a). 

As examples - using equation (16b), with $ G_{3} $ equal to the Coulomb potential or the Yukawa potential and substituting the resulting $ n^{2} $ in equation (16a) results in the corresponding relativistic time delayed potentials. This confirms that the steps in this section provides the correct prescription in deriving the relativistic two-body interaction potential. 

6. Given this relativistic two-body potential, we find that the effective action for the relativistic particles is now given by 
\begin{equation}\label{17}
 S_{eff}(\vec{x}^{a}) = S_{rp} - \frac{1}{2} \int dt dt'\sum_{a,b; a \neq b}^{N} \alpha^{2} G_{4}(\vert\vec{x}^{b}(t') -\vec{x}^{a}(t)\vert; t - t'),
 \end{equation}
 where we dropped the $ tr ln $ term in equation (11) as it is $ \vec{x}^{a} $ independent. It is also of the order $ \hslash $ thus could not be of the same contribution as the other terms in $ S_{eff} $. This action goes over to the non-relativistic many particle action for systems with two-body potential. And from this action, we derive the relativistic many particle Hamiltonian given by
 \begin{equation}\label{18}
H_{eff}(\vec{x}^{a}, \vec{p}^{a},t) = \sum_{a=1}^{N} (\vert\vec{p}^{a}\vert^{2} + m^{2})^{\frac{1}{2}}  + \frac{1}{2} \int dt'\sum_{a,b; a \neq b}^{N} \alpha^{2} G_{4}(\vert\vec{x}^{b}(t') -\vec{x}^{a}(t)\vert; t - t').
\end{equation}
This Hamiltonian should be the starting point of any statistical mechanics of many particle relativistic systems, the results of which can be directly compared to those of the non-relativistic system.

We can verify Poincare symmetry at the level of relativistic particles interacting with the scalar field. The action is given by $ S_{rp} + S_{s} + S_{int} $. We find the following generators of space-time symmetry operations
\begin{subequations}\label{19}
\begin{gather} 
 H =  \sum_{a=1}^{N}[(\vec{p_{a}})^{2}+m^{2}]^{\frac{1}{2}} + \int d^{3}x \textsf{H}(\vec{x},t) + \int d^{3}x \Phi(\vec{x},t) \sum_{a=1}^{N} \alpha \delta^{3}(\vec{x}+\vec{x}_{a}(t)),\label{first}\\
 P_{i} = \sum_{a=1}^{N} p_{i}^{a} + \int d^{3}x \Pi(\vec{x},t)\partial_{i}\Phi(\vec{x},t),\label{second}\\
 J_{i} = \epsilon_{ijk}\left[ \sum_{a=1}^{N} x_{j}^{a}p_{k}^{a} + \int d^{3}x \textsf{P}_{j}(\vec{x},t) x_{k}\right] ,\label{third}\\
 \begin{split}
 K_{i}& = \sum_{a=1}^{N} \left\lbrace \frac{1}{2}{x_{i}^{a}[(\vec{p_{a}})^{2}+m^{2}]^{\frac{1}{2}}+[(\vec{p_{a}})^{2}+m^{2}]^{\frac{1}{2}}x_{i}^{a}}\right\rbrace + \int d^{3}x \left\lbrace  x_{i} \textsf{H}(\vec{x},t) + x_{0} \textsf{P}_{i}(\vec{x},t) \right\rbrace\\
 &\quad  + \int d^{3}x \Phi(\vec{x},t)\sum_{a=1}^{N} \alpha \delta^{3}(\vec{x}+\vec{x}_{a}(t)) x_{i}.\label{fourth}
\end{split}
\end{gather}
\end{subequations}  
In verifying the Poincare algebra, we will see the crucial role of the positive sign in the definition of $ \textsf{j} $ given in equation (9b). Had it been negative, two similar terms would have added up and the algebra will not be satisfied. Thus at the level of the scalar-relativistic particles, the Poincare algebra is satisfied.

7. At the level of the effective dynamics of the relativistic particles defined by the effective action given in equation (17), we derive the generators of time translation, spatial translation, real rotations and boost. The other generators of the Poincare group in the effective theory aside from $ H_{eff} $ given in equation (18) are
\begin{subequations}\label{20}
\begin{gather}
\vec{P}_{eff} = \sum_{a=1}^{N} \vec{p}^{a},\label{first}\\
\vec{J}_{eff} = \sum_{a=1}^{N} \vec{x}^{a} X \vec{p}^{a},\label{second}\\
\begin{split}
\vec{K}_{eff}& = \frac{1}{2}\sum_{a=1}^{N} \{ \vec{x}^{a} [ (\vec{p_{a}}^{2}+m^{2})^{\frac{1}{2}} + \frac{1}{2} \int dt'\sum_{b \neq a}^{N} \alpha^{2} G_{4}(\vert\vec{x}^{b}(t') -\vec{x}^{a}(t)\vert; t - t')]\\
&\quad + [ (\vec{p_{a}}^{2}+m^{2})^{\frac{1}{2}} + \frac{1}{2} \int dt'\sum_{b \neq a}^{N} \alpha^{2} G_{4}(\vert\vec{x}^{b}(t') -\vec{x}^{a}(t)\vert; t - t')]\vec{x}^{a} \} \label{third}
\end{split}
\end{gather}
\end{subequations}
The generators for spatial translation and real rotations are the same as in the non-interacting  relativistic particles case. The generators for time translation and boosts are substantially altered by a non-local in time term.  It is this extra term, the time delayed two body interaction term that violates the Poisson brackets of the Poincare algebra even though each of the infinitesimal transformations is a symmetry.

Why is this so? The Poincare algebra involve equal time Poisson bracket relations, which is consistent with the one time formulation of special relativity used here, the isochronous frame. But the time delay in $ G_{4} $, shows that when we consider the effect of the other particles on say particle 1 with coordinate $ \vec{x}^{1}(t) $ at time t, not all particles b, with  b from 2 to N at earlier time t' will affect particle 1 at time t. Only those particles consistent with the time delay defined by $ G_{4} $ will. Other particles would have affected particle 1 maybe even at times earlier than t' while others would affect particle 1 at later than t'. In other words, we have lost the 'isochronocity' of the time delay in the interaction of the particles. It is for these reason that the Poincare algebra, which is an equal time Poisson bracket is no longer satisfied by the generators of the infinitesimal space-time transformations even if they are separately conserved. 

8. When we let $ c \rightarrow \infty $, $ G_{4} $ becomes the instantaneous two-body potential. The effective action given by equation (17) becomes
\begin{equation}\label{21}
S_{nr} = \int dt \left[ \frac{1}{2}\sum_{a=1}^{N} m \dot{\vec{x}}^{a} \cdot \dot{\vec{x}}^{a}  - \sum_{a, b, a \neq b}^{N} V(\vert \vec{x}^{a} - \vec{x}^{b} \vert ) \right] ,
\end{equation}
where we dropped the rest mass energy constant. This action is invariant under infinitesimal time translation, spatial translation, real rotation and boosts $ ( \vec{x}^{a} \rightarrow \vec{x}^{a} + \vec{\alpha} t ) $. The generators of these transformations are
\begin{subequations}\label{22}
\begin{gather}
H = \frac{1}{2} \sum_{a=1}^{N} \frac{1}{m} \vec{p}^{a} \cdot \vec{p}^{a} + \sum_{a, b, a \neq b}^{N} V(\vert \vec{x}^{a} - \vec{x}^{b} \vert), \label{first}\\
\vec{P} = \sum_{a=1}^{N} \vec{p}^{a}, \label{second}\\
J_{i} = \epsilon_{ijk} \sum_{a=1}^{N} x_{j}^{a} p_{k}^{a},\label{third}\\
\vec{K} = \sum_{a=1}^{N} (\vec{p}^{a} t - m\vec{x}^{a}).\label{fourth}
\end{gather}
\end{subequations}
It is simple to show that these generators satisfy the Galilean group, the $ c \rightarrow \infty $ contraction of the Poincare group, and it is defined by defined by 
\begin{subequations}\label{23}
\begin{gather}
[P_{i}, H] = [J_{i}, H] = [P_{i}, P_{j}] = [K_{i}, K_{j}] = 0,\label{first}\\
[J_{i}, J_{j}] =\epsilon_{ijk} J_{k}, \label{second}\\
[J_{i}, P_{j}] = \epsilon_{ijk} P_{k}, \label{third}\\
[J_{i}, K_{j}] = \epsilon_{ijk} K_{k}, \label{fourth}\\
[K_{i}, H] = P_{i}, \label{fifth}\\
[K_{i}, P_{j}] = \delta_{ij} m,\label{sixth}
\end{gather}
\end{subequations}

The above discussion and the discussions in number 2 (free relativistic particles) and number 6 (relativistic particles with local interaction with a scalar field, show that the generators of infinitesimal space-time symmetries satisfy the relevant algebras, Galilean for number 8 and Poincare for numbers 2 and 6. But number (7) clearly showed that when we derived the effective dynamics for the relativistic particles, the time delayed interaction term invalidated the algebra usually satisfied by the generators (Poincare in this case). But each of those generators generate the space-time transformations and are conserved. In particular the Hamiltonian includes a valid relativistic generalization of the two-body interaction potential. For this reason, the Hamiltonian derived here is relativistic and can be used as a starting point in a relativistic statistical mechanics with results that can be compared directly with the corresponding non-relativistic case. 

9. To summarize, suppose we have a non-relativistic Hamiltonian given by equation (22a). How do we derive the relativistic Hamiltonian? First, we identify the non-relativistic two-body potential $ V(\vert \vec{x}^{a} - \vec{x}^{b} \vert) $ with $ G_{3} $ of number 5. Then we compute $ n^{2}(\vert \vec{k} \vert^{2}) $ as given in equation (16b). Then make this function relativistic by extending the argument to $ n^{2}(\vert \vec{k} \vert^{2} - \frac{1}{c^{2}} k_{0}^{2}) $ and substitute this in equation (16a) to derive the time delayed two-body interaction potential. The relativistic Hamiltonian is then given by equation (18).

With this Hamiltonian, we can now do statistical mechanics, which hopefully will help settle the confusing relativistic thermodynamics results. Or, we can now work out the relativistic corrections to non-relativistic statistical mechanics of known systems. 

\begin{acknowledgments}
I would like to thank Ian Vega of the NIP for tutoring me on how to use Latex in Mac. He is also grateful to Felicia Magpantay for sending him pdf files of relevant papers and correcting the error in his latex file.
\end{acknowledgments}



\end{document}